\documentclass[aps,prd,superscriptaddress,twocolumn, showpacs, showkeys, amsmath, amssymb, floatfix]{revtex4-1}

\usepackage[normal]{subfigure}
\usepackage{color}
\usepackage{lipsum}
\usepackage{graphicx}

\usepackage{xspace} 
\usepackage{dcolumn}
\usepackage{braket}
\usepackage{bm}
\usepackage[normalem]{ulem}
\newcommand{\ambe}{${}^{241}$AmBe\xspace} 
\newcommand{\co}{${}^{60}$Co\xspace} 
\newcommand{\tor}{${}^{232}$Th\xspace} 
\newcommand{\coleff}{${}^{57}$Co\xspace}

\newcommand{\bern}{\affiliation{Albert Einstein Center for Fundamental Physics, University of Bern, 3012 Bern, Switzerland}}
\newcommand{\bologna}{\affiliation{Department of Physics and Astrophysics, University of Bologna and INFN-Bologna, 40126 Bologna, Italy}}
\newcommand{\coimbra}{\affiliation{Department of Physics, University of Coimbra, 3004-516, Coimbra, Portugal}}
\newcommand{\columbia}{\affiliation{Physics Department, Columbia University, New York, New York, 10027, USA}}
\newcommand{\lngs}{\affiliation{INFN-Laboratori Nazionali del Gran Sasso and Gran Sasso Science Institute, 67100 L'Aquila, Italy}}
\newcommand{\mainz}{\affiliation{Institut f\"ur Physik \& Exzellenzcluster PRISMA, Johannes Gutenberg-Universit\"at Mainz, 55099 Mainz, Germany}}
\newcommand{\heidelberg}{\affiliation{Max-Planck-Institut f\"ur Kernphysik, 69117 Heidelberg, Germany}}
\newcommand{\munster}{\affiliation{Institut f\"ur Kernphysik, Wilhelms-Universit\"at M\"unster, 48149 M\"unster, Germany}}
\newcommand{\nikhef}{\affiliation{Nikhef and the University of Amsterdam, Science Park, 1098XG Amsterdam, Netherlands}}
\newcommand{\nyuad}{\affiliation{New York University Abu Dhabi, Abu Dhabi, United Arab Emirates}}
\newcommand{\purdue}{\affiliation{Department of Physics and Astronomy, Purdue University, West Lafayette, Indiana 47907, USA}}
\newcommand{\rensselaer}{\affiliation{Department of Physics, Applied Physics and Astronomy, Rensselaer Polytechnic Institute, Troy, New York 12180, USA}}
\newcommand{\rice}{\affiliation{Department of Physics and Astronomy, Rice University, Houston, Texas 77005, USA}}

\newcommand{\stockholm}{\affiliation{Department of Physics, Stockholm University, Stockholm, SE-10691 Sweden}}
\newcommand{\subatech}{\affiliation{SUBATECH, Ecole des Mines de Nantes, CNRS/In2p3, Universit\'e de Nantes, Nantes 44307, France}}
\newcommand{\torino}{\affiliation{INFN-Torino and Osservatorio Astrofisico di Torino, 10125 Torino, Italy}}
\newcommand{\ucla}{\affiliation{Physics \& Astronomy Department, University of California, Los Angeles, California, 90095, USA}}
\newcommand{\weizmann}{\affiliation{Department of Particle Physics and Astrophysics, Weizmann Institute of Science, Rehovot, 7610001, Israel}}
\newcommand{\zurich}{\affiliation{Physik-Institut, University of Zurich, 8057 Zurich, Switzerland}}

\begin{document}
\preprint{APS/123-QED}

\title{Low-mass dark matter search using ionization signals in XENON100}

\author{E.~Aprile}\columbia
\author{J.~Aalbers}\nikhef
\author{F.~Agostini}\lngs\bologna
\author{M.~Alfonsi}\mainz
\author{F.~D.~Amaro}\coimbra
\author{M.~Anthony}\columbia
\author{F.~Arneodo}\nyuad
\author{P.~Barrow}\zurich
\author{L.~Baudis}\zurich
\author{B.~Bauermeister}\mainz\stockholm
% ...

\date{\today} 

\email{\{atiseni,ctunnell\}@nikhef.nl; ran.budnik@weizmann.ac.il}
\collaboration{XENON Collaboration}

\begin{abstract}
We perform a low-mass dark matter search using an exposure of 30\,kg$\times$yr with the XENON100 detector. By dropping the requirement of a scintillation signal and using only the ionization signal to determine the interaction energy, we lowered the energy threshold for detection to 0.7\,keV for nuclear recoils. No dark matter detection can be claimed because a complete background model cannot be constructed without a primary scintillation signal. Instead, we compute an upper limit on the WIMP-nucleon scattering cross section under the assumption that every event passing our selection criteria could be a signal event. Using an energy interval from 0.7\,keV to 9.1\,keV, we derive a limit on the spin-independent WIMP-nucleon cross section that excludes WIMPs with a mass of 6\,GeV/$c^2$ above $1.4 \times 10^{-41}$\,cm$^2$ at 90\% confidence level.
\end{abstract}

\maketitle

\section{\label{sec:intro}Introduction}

Astrophysical observations indicate that dark matter (DM) is needed to explain structures ranging from the scales of galaxies to the largest observed scales~\cite{darktheory}.  Nevertheless, little is known about its nature.  One theoretically favored candidate is a weakly interacting massive particle (WIMP). These particles may be detectable with experiments sensitive to WIMP-induced nuclear recoils~\cite{wimp}.

Most WIMP models predict particles with a mass at the electroweak scale of $\sim$100\,GeV/$c^2$~\cite{100gevwimp}.  However, there is also interest in light-mass DM, below 10\,GeV/$c^2$, prompted by, e.g., asymmetric models~\cite{asymmetricmodels, asymmetricmodels1} and claims of DM observations~\cite{dama,cdmssi}. 
Light-mass DM would yield low-energy events that are close to the experimental energy threshold of liquid-xenon detectors. Therefore, exploiting an approach that lowers the threshold~\cite{XENON10}, we investigate the spin-independent WIMP-nucleon cross section versus mass parameter space extending the XENON100 results for masses below $\sim$7.4\,GeV/$c^2$.

\section{\label{sec:xenon100} The XENON100 detector}

The XENON100 detector~\cite{detector} is a dual-phase (liquid-gas) xenon time projection chamber (TPC) located in the Laboratori Nazionali del Gran Sasso (LNGS). The TPC detection principle allows for measurements of nuclear recoils (NR) and electronic recoils (ER) through two signals: a prompt scintillation signal S1 and an ionization signal S2. The S1 signal is scintillation light from the rapid deexcitation of excited liquid xenon molecular states after an ionizing particle deposits energy. This deposition also liberates electrons, which drift in an electric field of 530\,V/cm toward the liquid-gas interface, where a larger  field of $\sim$12\,kV/cm extracts them from the liquid. These accelerated electrons generate proportional scintillation in the xenon gas above the liquid. 

Two arrays of 178 1''-square Hamamatsu R8520-AL PMTs are installed above and below the 62-kg xenon target. They detect both signals from the target.  The distribution of the S2 signal among the top PMTs gives the projection of the interaction position on the PMT plane, while the relative time between the S1 and S2 signals provides the depth of the interaction, or $z$ coordinate. We distinguish ER and NR by the ratio of their respective S1 and S2 signals. A trigger identifies S2 signals, and the waveform of each PMT is digitized in the interval between 200\,$\mu$s before and after the trigger. The time for an electron to drift from the cathode to anode, or the maximum drift time, is
176\,$\mu$s~\cite{detector}. The TPC is surrounded by an active veto region consisting of 99\,kg of liquid xenon, instrumented with 64 PMTs optically isolated from the TPC. 

In previous XENON100 analyses~\cite{225livedays,spindependent}, the recoil energy has been determined using the size of the S1 signal and the relative scintillation efficiency for the nuclear recoils, $\mathcal{L}_{\textrm{eff}}$, relative to the 122\,keV calibration $\gamma$ line of \coleff~\cite{225livedays}. WIMPs with a mass below 10\,GeV/$c^2$ create NRs only up to a few keV, resulting in an S2 signal lower than a few hundred photoelectrons (PE) and an S1 signal that is often not detectable.  Therefore for this analysis we only use the S2 signal to infer the energy.

\section{\label{sec:data}Analysis}

This analysis is performed using the data from XENON100's Science Run II, which collected a 225 live-day exposure between February 28, 2011 and March 31, 2012~\cite{225livedays}.
For the WIMP analysis, we drop the requirement of observing an S1 signal. This allows us to lower the effective threshold at the cost of losing $z$ coordinate reconstruction from the S2-S1 peak time difference and particle identification based on the S2/S1 signal ratio.  We perform a background-limited analysis on this previously unblinded data set.  

Both a NR and an ER within liquid xenon will produce an S2 signal. We use calibration data of ERs and NRs taken with external \co/\tor and \ambe calibration sources, respectively. In these calibrations and in the DM search data, photo-ionization and delayed extraction of electrons produce signals that have a mean size of 20\,PE per electron~\cite{singleelectron}. We restrict ourselves to charge signals above 80\,PE, where the trigger efficiency is still at 80\%, to minimize the background from these electrons. For the same reason, this value will be used as the lower threshold for the WIMP analysis. 

Many processes besides WIMP interactions can create S2 or S2-like signals in our detector, e.g., radioactive backgrounds or photo-ionization of impurities or metallic surfaces in the TPC~\cite{singleelectron}. We use selection criteria to suppress these backgrounds in the DM search data. To begin, WIMPs are expected to interact uniformly in the liquid xenon target. In the DM search data, the event rate increases towards the radial edges of the detector because of radioactive backgrounds. Therefore, we require that the reconstructed radius of the event is less than 13.4\,cm, which is approximately 2\,cm from the TPC walls. This cut removes events from external backgrounds, which are stopped predominantly in the outer layers of the liquid target. The remaining liquid xenon target mass is 48.3\,kg \cite{analysispaper}.  Within this target, the events are uniformly distributed radially, which means that a smaller fiducial volume does not reduce the background density. 

Given an event with an S1 signal in the DM search data, we can use the information from that signal to isolate nuclear recoils using two additional cuts. First, the Monte Carlo nuclear-recoil model of~\cite{MarcWeber} is used to determine a cut on the S1 size relative to the S2 size for any WIMP mass less than 20\,GeV/$c^2$. We parametrize this Monte Carlo model by requiring that a nuclear recoil has--if present--an S1 signal less than $[4.7 + 0.012 \times (\text{S2} - 80)]$\,PE. This cut has an acceptance of 99.9\% determined from the same Monte Carlo. Second, we can estimate the $z$ position of the interaction from the drift time between the S1 and S2 signals. We require the $z$ position to be more than 1.9\,cm below the liquid-gas interface and more than 0.5\,cm above the cathode. This condition decreases the fiducial volume by 8\% and we conservatively assume an acceptance of 92\% also for events without a detected S1 signal.

\begin{figure}
\centering
\includegraphics[]{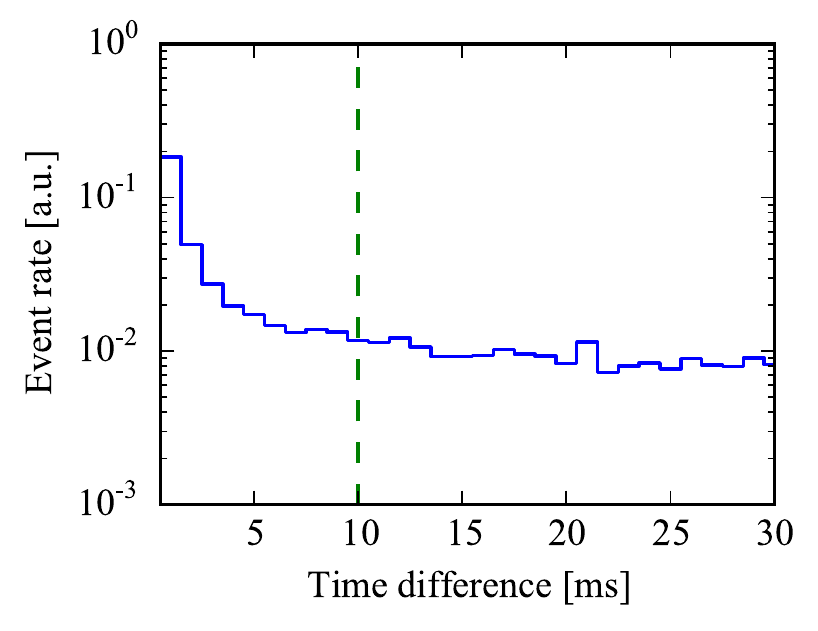}
\caption{Rate of events (S2 $>$ 80\,PE) as a function of the time difference from the previous recorded event. A cut is set at 10\,ms to remove a population of events of small S2 signals (e.g., photo-ionization) that appears within a few ms from the previous trigger.}
\label{fig:triggerhold}
\end{figure}

Secondary S2 signals can create events in which the main S2 signal is preceded or followed by similar nearby signals in the same event. These can be caused by multiple scatters in the active volume (i.e., not WIMPs) or misidentified detector artifacts. Additionally, any interaction in XENON100 can cause small S2 signals appearing up to milliseconds after the trigger, which are partly caused by photo-ionization on metal surfaces or impurities and possibly by  delayed charge extraction as well~\cite{singleelectron}. 

We remove events which occur less than 10\,ms after any other recorded event, resulting in a 2\% live-time reduction. Figure~\ref{fig:triggerhold} shows the event rate as a function of the time difference from the previous event. Signals caused by photo-ionization or delayed extraction are observed within a few ms from the previous event and are removed by this selection.

\begin{figure}
\centering
\includegraphics[]{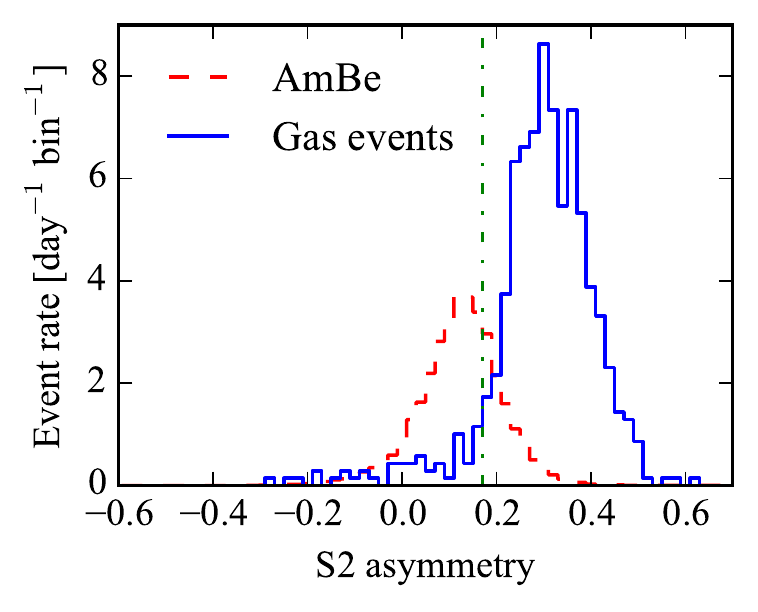}
\caption{S2 asymmetry parameter for \ambe calibration data in the liquid and a population of events produced in the xenon gas phase. We select interactions in the gas by requiring an S1 signal, small drift time, and a large S2 width using \co and DM search data.  An S2 asymmetry cut set at 0.17 is used to reject the gas event population in the dark matter data.}
\label{fig:asymmetrydist}
\end{figure}

In the DM search data, we reject events with more than one S2 signal in the same event, e.g., multiple scatter events.  If an S2 signal larger than 10 (30)\,PE is seen 176\,$\mu$s before (after) the main S2 signal, the event is removed. The threshold after the main S2 signal is less strict since even a 250\,PE S2 signal will itself create a secondary single-electron S2 signal ($\approx$ 20\,PE~\cite{singleelectron}) by photo-ionization in 10\% of the cases. The acceptance loss is 3\% at S2 = 100\,PE and slowly increasing for larger S2 signals, as estimated by a model of induced S2 signals similar to~\cite{singleelectron}, but extended to low energies using the 10\,ms time difference cut. 

For the following selection criteria, we estimate the acceptance on calibration data. For calibration events, to ensure that we only select valid low-energy events, we additionally require that the S1 signal observed in the TPC (at any size) has a coincident S1 signal in the veto region. We also apply the fiducial volume and single scatter selections as described above. In this way, we create a low-energy sample of real interactions. We use the fraction of events removed by the individual selection condition in the \ambe calibration data~\cite{analysispaper}. \ambe calibration was acquired before and after  the DM search data. The acceptance for \ambe taken at the end of the run is $\simeq$~6\% lower compared to \ambe acquired at the beginning of the run. Conservatively, we choose \ambe calibration data acquired at the end of the run to model the WIMP acceptance.

Events which contain too much electronic noise activity cannot be evaluated properly and are removed by comparing the area of the main S2~peak to the remaining baseline area. An S2 size-dependent threshold (the S2 pulse should contain at least 45\% of the total area at 100\,PE) was derived using \ambe calibration data and leads to a 97\% WIMP acceptance.

\begin{table}[!h]
\caption{Acceptances of the different data selections and number of DM candidate events passing the selections. The cuts are applied sequentially. The number of events is in the S2 energy range [80, 1000]\,PE.}\label{table1}
\begin{tabular}{ l | m{2.1cm} | m{2.4cm}}
Description of cut & Acceptance at S2=100\,PE  & \multicolumn{1}{r|}{Events} \\
\hline 
Radial cut (starting events)  & \multicolumn{1}{r|}{100\%} &  \multicolumn{1}{r|} {254901}  \\ 
Depth and electronic recoil &  \multicolumn{1}{r|}{92\%} & \multicolumn{1}{r|}{103914} \\ 
Detector noise & \multicolumn{1}{r|}{97\%}  & \multicolumn{1}{r|}{57516} \\
Single S2 and 10\,ms cut & \multicolumn{1}{r|}{95\%}  & \multicolumn{1}{r|}{49041} \\
Interaction in the gas & \multicolumn{1}{r|}{61\%} &  \multicolumn{1}{r|}{13560} \\

\end{tabular}
\end{table}

Finally, we apply a cut to remove events where the S2 signal is produced by an interaction in the gas between the anode and the top PMT screening electrode~\cite{detector}. These are most likely caused by radioactivity from the top PMT array. In these so-called ``gas events,'' a larger than average fraction of the S2 light is seen by the top PMT array since the S2 signal is produced close to it. 
The S2 signal is also wider than an S2 produced in the liquid since the luminescence region is typically twice as wide and--if an S1 signal is detected--it occurs very shortly before the S2. Therefore, we define an asymmetry parameter (S2$_{\text{top}}$ $-$ S2$_{\text{bottom}}$) / (S2$_{\text{top}}$ $+$ S2$_{\text{bottom}}$), corresponding to the fraction of observed light in the top PMTs compared to the bottom PMTs. 

In Fig.~\ref{fig:asymmetrydist}, the asymmetry parameter is shown for \ambe events that occurred in the liquid xenon and a sample of events from interactions in the gas phase.
The gas events are taken from \co and DM search data, requiring an S1 signal  and selecting events where the S2 width at 10\% peak height is inconsistent with diffusion broadening given the drift time of the event. Both distributions are normalized to the rate expected in the DM search data. The events in the liquid should be primarily due to ERs from background $\gamma$s, so we estimate the rate by comparing the rate of \co events and DM search data events at energies far beyond the region of interest, as done in~\cite{analysispaper}. The gas event rate was estimated from DM search data events with an S2 asymmetry larger than 0.45 (again, well beyond the region of interest), as seen in Fig.~\ref{fig:asymmetrydist}.

\begin{figure}
\centering
\includegraphics[]{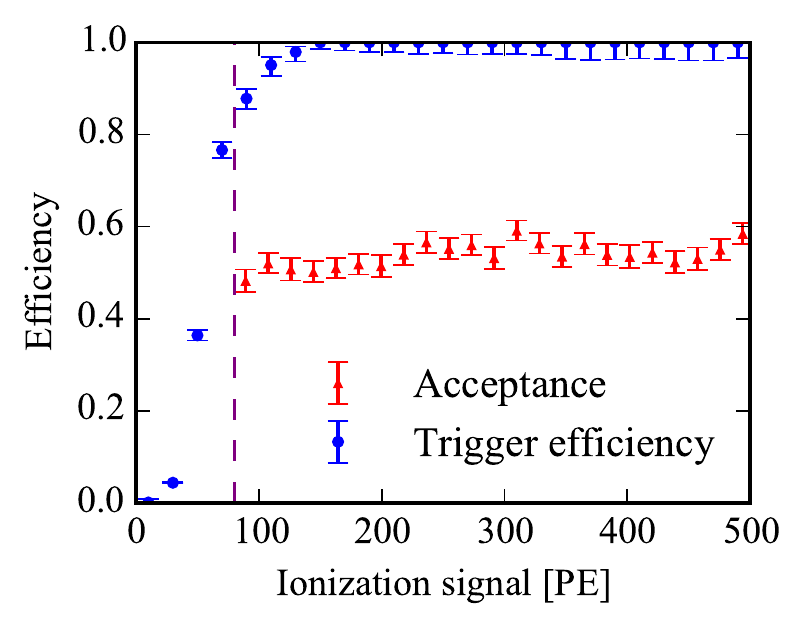}
\caption{\label{fig:total_acceptance_and_trigger} The analysis acceptance (red triangles) and the trigger efficiency (blue circles). The purple dashed line is the analysis threshold (80\,PE).}
\end{figure}
We remove events with an S2 asymmetry parameter larger than 0.17 and smaller than an S2 size-dependent threshold derived from \ambe ($-$0.32 at 100\,PE). The 0.17 threshold is chosen by optimizing the ratio of the liquid events over the square root of gas events. Only 61\% of liquid events with an S2 signal of 100\,PE will pass the asymmetry cut (as determined from the \ambe data). The low acceptance is necessary because of the gas event background in this analysis. 
We also apply a loose S2 10\%-width selection of [0.8, 2.7]\,$\mu$s with an acceptance of 99.8\% at S2=100\,PE.

\begin{figure}[!h]
\centering
\includegraphics[]{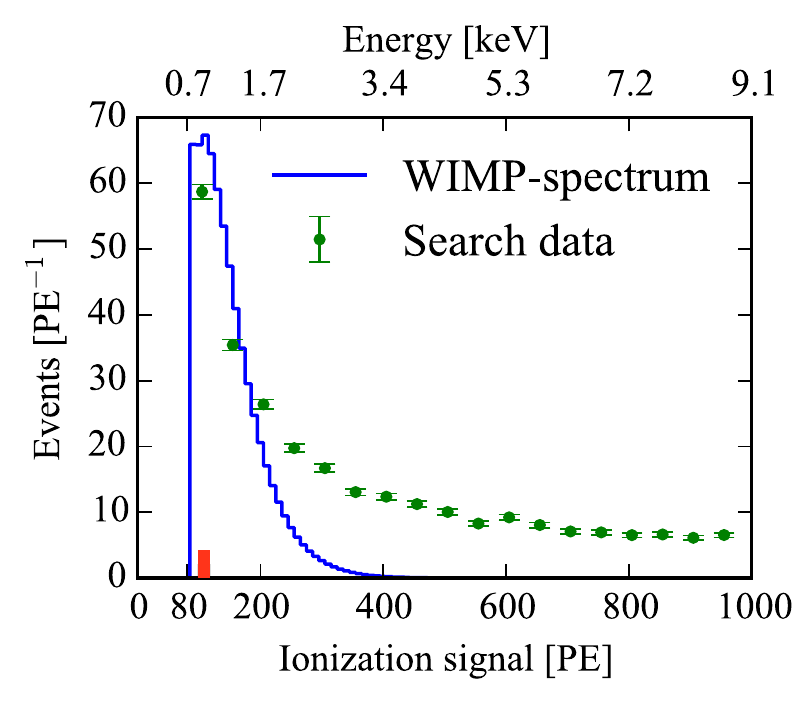}
\caption{\label{fig:rate} Energy distribution of the events remaining in the data set after all data selection cuts. As an example, the expected spectrum for a WIMP of 6\,GeV/$c^2$ and a spin-independent WIMP-nucleon scattering cross section of $1.5 \times 10^{-41}$\,cm$^2$ is also shown. The corresponding nuclear recoil energy scale is indicated on the top axis. The charge yield model assumed here has a cutoff at 0.7\,keV, which truncates the WIMP spectrum. The optimum interval (thick red line) is found in the S2 range [98, 119]\,PE and contains 1173 events.}
\end{figure}

Figure~\ref{fig:total_acceptance_and_trigger} shows the analysis acceptance and the trigger efficiency \cite{analysispaper} as a function of the S2 signal size.  The trigger efficiency in our region of interest is more than 80\%.  The product of the trigger efficiency and analysis acceptance is our final signal detection efficiency. Table~\ref{table1} shows the acceptance of the analysis selections discussed above, as well as the number of events remaining at each stage. After applying the data selection cuts summarized in Table \ref{table1} to the the entire data set of 30\,kg\,$\times$\,yr, 13560 valid candidate events remain in the S2 range [80, 1000]\,PE (see Fig.~\ref{fig:rate}).

\section{Results}\label{sec:result}

\begin{figure}
\centering
\includegraphics[]{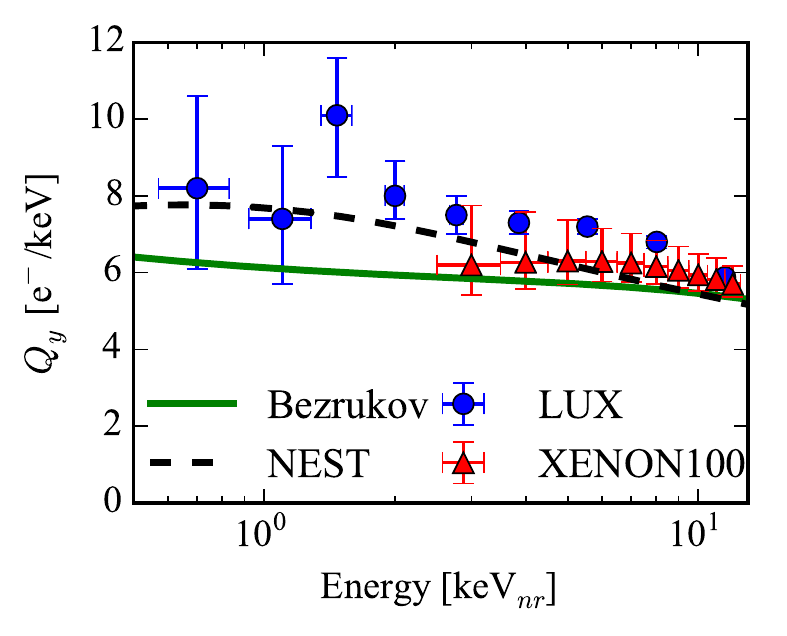}
\caption{\label{fig:qy} Charge yield ($Q_y$) as a function of energy for nuclear recoils (keV). This analysis employs the conservative nuclear recoil charge yield  model of Bezrukov \emph{et al.} (electric field independent)~\cite{bezrukov}, given by the green line. It agrees with the measurement of XENON100 (E = 0.53\,kV/cm)~\cite{MarcWeber} (red triangles). The NEST model (E = 0.73\,kV/cm)~\cite{nest} (dashed black) and the recent measurement of LUX (E = 0.18\,kV/cm)~\cite{luxlowenergy} (blue points) predict slightly higher yields. To account for the mild discrepancies below 3\,keV, we use the model from Bezrukov \emph{et al.} and conservatively assume $Q_y$=0 below 0.7\,keV.}
\end{figure}

The interpretation of the outcome of the data selection requires the reconstruction of a nuclear recoil equivalent energy scale from the measured S2 signals. It is based on two quantities: the first one is the charge yield $Q_y$, shown in Fig.~\ref{fig:qy}, which gives the number of ionization electrons per keV liberated by a NR event. The second one is the secondary scintillation gain $Y$, which is detector-dependent and gives  the number of proportional scintillation photoelectrons per electron extracted into the gas phase. In this science run of XENON100, $Y$ is described by a normal distribution with $\mu = (19.7 \pm 0.3)$\,PE/e$^-$ and $\sigma = (6.9 \pm 0.3)$\,PE/e$^-$~\cite{singleelectron}. Charge extraction from the liquid is almost unity at the XENON100 extraction field~\cite{detector}.

As shown in Fig.~\ref{fig:qy}, there is some remaining uncertainty in $Q_y$, especially at very low recoil energies, even though the LUX data demonstrate clearly that $Q_y$ is nonzero above 0.7\,keV~\cite{luxlowenergy}. In order to not base our WIMP result on optimistic assumptions, we use the analytical model of Bezrukov \emph{et al.}~\cite{bezrukov}, which agrees with the XENON100 measurement~\cite{MarcWeber}, and the NEST model~\cite{nest} above $\sim$6\,keV and is more conservative at lower energies. We additionally introduce a cutoff at 0.7\,keV, below which $Q_y$ is set to zero, to penalize the result for the limited knowledge on the charge yield at the lowest energies. This energy also corresponds to the threshold at which signals will be above our 80\,PE threshold.

\begin{figure}[!h]
\centering
\includegraphics[]{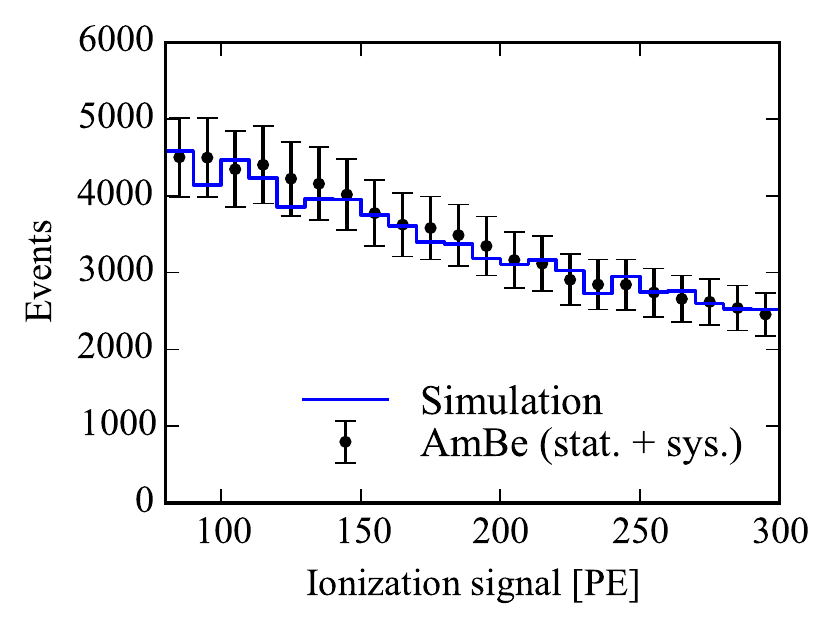}
\caption{\label{fig:enr_threshold} S2 spectrum of \ambe calibration data compared to simulations using the $Q_{y}$ from Bezrukov \emph{et al.}~\cite{bezrukov} with no energy cutoff.}
\end{figure}

However, we note that a Monte Carlo model based on the Bezrukov \emph{et al.} function without any cutoff leads to a good description of the measured charge spectrum from \ambe calibration data (see Fig.~\ref{fig:enr_threshold}).  The data were selected based on the same criteria as used in the WIMP analysis, with the exception of the S2~asymmetry cut, which is not required due to the significantly higher rate of the \ambe source compared to the gas event rate. Besides the statistical uncertainty, the spectrum also includes a systematic uncertainty of 8\%, which is mainly due to the uncertainties in the S2 amplification~\cite{singleelectron} and the cut acceptance. The simulation follows the strategy described in~\cite{MarcWeber} but ignores the S1 light information. 

% Below are the data points for this result in the units indicated in the axis.
%mass=[3.0,3.40446210263,3.86345406942,4.38432765487,4.97542578217,5.64621617329, 6.40744299507, 7.2712989505, 8.2516205713, 9.36410984009, 10.6265856918, 12.0592694227, 13.6851085784, 15.5301445085, 17.6239294759, 20.0]
%xsec=[6.534788902911208e-39, 7.2532759897689746e-40, 1.7603412454922315e-40, 6.353077392458506e-41, 2.9634932978431705e-41, 1.6451008447868074e-41, 1.0363095557612012e-41, 7.23470690825168e-42, 5.508795283069165e-42, 4.516243593921473e-42, 3.880680688857027e-42, 3.3947126677656e-42, 2.9649500044730256e-42, 2.6245924773640006e-42, 2.3991307721031054e-42, 2.24203741440827e-42]

\begin{figure}
\centering
\includegraphics[height=6.5cm]{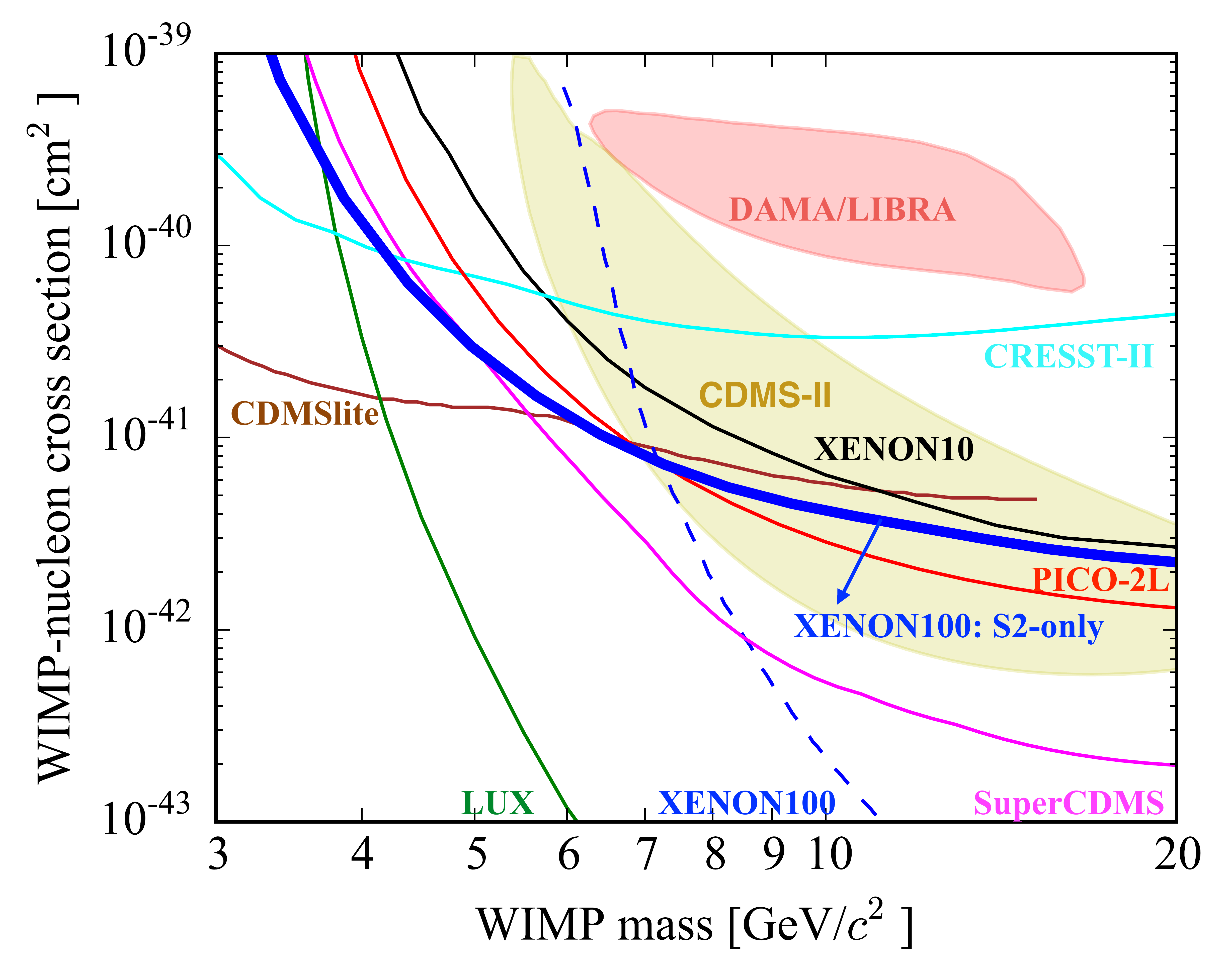}
\caption{\label{fig:limit} WIMP exclusion limit on the spin-independent WIMP-nucleon scattering cross section at 90\% confidence level. Limits from the LUX~\cite{lux}, XENON100~\cite{225livedays}, SuperCDMS~\cite{supercdms}, CDMSlite~\cite{cdmslite}, XENON10~\cite{XENON10}, CRESST-II~\cite{CREEST2015} and PICO-2L~\cite{PICO2016} experiments are shown. The claims from DAMA/LIBRA experimental data~\cite{damachanneling} and CDMS-II (Si detectors)~\cite{cdmssi} are also shown. The limit from this analysis is shown with the thick blue line and it improves the XENON100 result~\cite{225livedays} (dashed blue line) for WIMP masses below $\sim$7.4\,GeV/$c^2$.}
\end{figure}

The same Monte Carlo method is used to model the expected WIMP energy spectra. The number of electrons released after a nuclear recoil of energy $E$ is given by a Poisson distribution with mean $N = EQ_y$. 
The charge loss due to the electron lifetime ($\tau_e$) is modeled per event as an exponential reduction in the number of electrons, though this effect is small due to the average $\braket{\tau_e}  = 570$\,$\mu$s.  The evolution of $\tau_e$ throughout the 225 days is modeled as in previous work \cite{analysispaper}.
The secondary scintillation is modeled using the measured parameters given above and in~\cite{singleelectron}. A Maxwell-Boltzmann distribution with the asymptotic velocity of the local system $v_0=220$\,km/s, the solar velocity $v_{sun}=232$\,km/s and the galactic escape velocity $v_{esc}=544$\,km/s is used to model the DM halo, assuming a local WIMP density of $\rho_0=0.3$\,GeV/($c^2$$\times$cm$^3$)~\cite{green}. As an example, Fig.~\ref{fig:rate} shows the NR spectrum, as parametrized in~\cite{lewinsmith},  induced by a 6\,GeV/$c^2$ WIMP at a spin-independent cross section of $\sigma=1.5 \times 10^{-41}$\,cm$^2$. We observe an event rate of $\sim$0.5\,events/(keV$\times$kg$\times$day) between 0.7 and 1.7\,keV that drops to $\sim$0.07\,events/(keV$\times$kg$\times$day) between 3.4 and 9.1\,keV.

In the absence of a full background model, which cannot be constructed as the origin of the small-S2 background in the detector cannot be reliably quantified, we assume that every event passing the analysis cuts could be due to a DM interaction. The analysis employs the optimum interval method~\cite{optinterval} and will therefore always lead to an exclusion limit. The optimum S2 interval varies with WIMP mass, but
in all cases in this analysis, it contains a minimum of 1000 events
passing all cuts.
The low-mass WIMP result for this 30\,kg\,$\times$\,yr XENON100 exposure is based on all events remaining in the 80-1000\,PE interval (0.7-9.1\,keV), the NR acceptance of Fig.~\ref{fig:total_acceptance_and_trigger}, and is shown in Fig.~\ref{fig:limit}. At a WIMP mass of 6\,GeV/$c^2$, XENON100 excludes spin-independent WIMP-nucleon interaction cross sections of $1.4 \times 10^{-41}$\,cm$^2$ at 90\% confidence level. The moderate improvement upon the XENON10 low-mass result~\cite{XENON10}, despite the much larger exposure, is due to the significantly higher background from photo-ionization events, which is enhanced by the presence of larger metal surfaces inside the TPC. 
The new result challenges a standard WIMP interpretation of the DAMA/LIBRA modulation signal, excludes large fractions of the CDMS-II (Si) preferred region and improves the result of the previous XENON100 result~\cite{225livedays} below $\sim$7.4\,GeV/$c^2$. We improve the LUX~\cite{lux} (SuperCDMS~\cite{supercdms}) results below $\sim$3.7 (5.3)\,GeV/$c^2$.

\section{\label{sec:acknowledgments} Acknowledgments}

We gratefully acknowledge support from the National Science Foundation, Swiss National
Science Foundation, Deutsche Forschungsgemeinschaft, Max Planck Gesellschaft,
Foundation for Fundamental Research on Matter, Weizmann Institute of Science, I-CORE,
Initial Training Network Invisibles (Marie Curie Actions, PITN- GA-2011-289442), Fundacao
para a Ciencia e a Tecnologia, Region des Pays de la Loire, Knut and Alice Wallenberg
Foundation, and Istituto Nazionale di Fisica Nucleare. We are grateful to Laboratori Nazionali del Gran Sasso for hosting and supporting the XENON project.

\end{document}